\documentclass[letterpaper]{article} 
\usepackage{aaai25}  
\usepackage{times}  
\usepackage{helvet}  
\usepackage{courier}  
\usepackage[hyphens]{url}  
\usepackage{graphicx} 
\usepackage{natbib}  
\usepackage{caption} 
\usepackage{soul}
\usepackage{graphicx}
\usepackage{booktabs}
\usepackage{float}
\usepackage{overpic}
\usepackage{subfloat}
\usepackage[skip=0.5\baselineskip]{caption}
\usepackage{bm}
\usepackage{amsmath}
\usepackage{booktabs}
\usepackage{multirow}
\usepackage{multicol}
\usepackage{enumitem}
\usepackage{subcaption}
\usepackage{threeparttable}
\usepackage{xspace}
\usepackage{amssymb}
\usepackage{color}
\usepackage{algorithmicx,algorithm}
\usepackage{algpseudocode}
\usepackage{algorithm,algpseudocode}
\usepackage{marvosym}   
\usepackage{colortbl}   
\usepackage{tcolorbox}  
\tcbuselibrary{breakable}   
\usepackage{wasysym}    
\usepackage[figuresright]{rotating} 

\usepackage{newfloat}
\usepackage{listings}
\DeclareCaptionStyle{ruled}{labelfont=normalfont,labelsep=colon,strut=off} 
\floatstyle{ruled}
\newfloat{listing}{tb}{lst}{}
\floatname{listing}{Listing}
\usepackage{xspace}

\newcommand{\eg}{\emph{e.g.,}\xspace}
\newcommand{\ie}{\emph{i.e.,}\xspace}

\title{LLMEmb: Large Language Model Can Be a Good Embedding Generator \\ for Sequential Recommendation}
\author {
    Qidong Liu\textsuperscript{\rm 1, 2},
    Xian Wu\textsuperscript{\rm 3} \thanks{Corresponding authors: Xian Wu, Xiangyu Zhao, Feng Tian},
    Wanyu Wang\textsuperscript{\rm 2},
    Yejing Wang\textsuperscript{\rm 2},
    Yuanshao Zhu\textsuperscript{\rm 2}, \\ 
    Xiangyu Zhao\textsuperscript{\rm 2 \ *}, 
    Feng Tian\textsuperscript{\rm 4 \ *}, 
    Yefeng Zheng\textsuperscript{\rm 3, 5}
}
\affiliations {
    \textsuperscript{\rm 1}School of Auto. Science \& Engineering, MOEKLINNS Lab, Xi'an Jiaotong University\\
    \textsuperscript{\rm 2}City University of Hong Kong \\
    \textsuperscript{\rm 3}Jarvis Research Center, Tencent YouTu Lab \\
    \textsuperscript{\rm 4}School of Comp. Science \& Technology, MOEKLINNS Lab, Xi’an Jiaotong University\\
    \textsuperscript{\rm 5}Medical Artificial Intelligence Lab, Westlake University\\
    liuqidong@stu.xjtu.edu.com, \{kevinxwu, yefengzheng\}@tencent.com, \{wanyuwang4-c, yejing.wang\}@my.cityu.edu.hk \\
    yuanshao@ieee.org, xianzhao@cityu.edu.hk, fengtian@mail.xjtu.edu.cn
}

\usepackage{bibentry}

\begin{document}

\maketitle

\begin{abstract}

    Sequential Recommender Systems (SRS), which model a user's interaction history to predict the next item of interest, are widely used in various applications. However, existing SRS often struggle with low-popularity items, a challenge known as the long-tail problem. This issue leads to reduced serendipity for users and diminished profits for sellers, ultimately harming the overall system. Large Language Model (LLM) has the ability to capture semantic relationships between items, independent of their popularity, making it a promising solution to this problem.
    In this paper, we introduce \textbf{LLMEmb}, a novel method leveraging LLM to generate item embeddings that enhance SRS performance. To bridge the gap between general-purpose LLM and the recommendation domain, we propose a Supervised Contrastive Fine-Tuning (SCFT) approach. This approach includes attribute-level data augmentation and a tailored contrastive loss to make LLM more recommendation-friendly. Additionally, we emphasize the importance of integrating collaborative signals into LLM-generated embeddings, for which we propose Recommendation Adaptation Training (RAT). This further refines the embeddings for optimal use in SRS.
    The LLMEmb-derived embeddings can be seamlessly integrated with any SRS models, underscoring the practical value. Comprehensive experiments conducted on three real-world datasets demonstrate that LLMEmb significantly outperforms existing methods across multiple SRS models. 
    The code for our method is released online \textcolor{blue}{https://github.com/Applied-Machine-Learning-Lab/LLMEmb}.
    
\end{abstract}


\section{Introduction}

Sequential recommender systems (SRS) have been extensively applied across various practical scenarios, such as e-commerce~\cite{zhou2018deep} and short video~\cite{pan2023understanding}.
The primary objective of SRS is to capture users' preferences based on their historical interactions and predict the next most possible item~\cite{fang2020deep}. 
To achieve this, many research studies have committed to developing neural network architectures for better modeling user interaction history, \eg SASRec~\cite{kang2018self} and Bert4Rec~\cite{sun2019bert4rec}.

\begin{figure}[!t]
\centering
\includegraphics[width=1\linewidth]{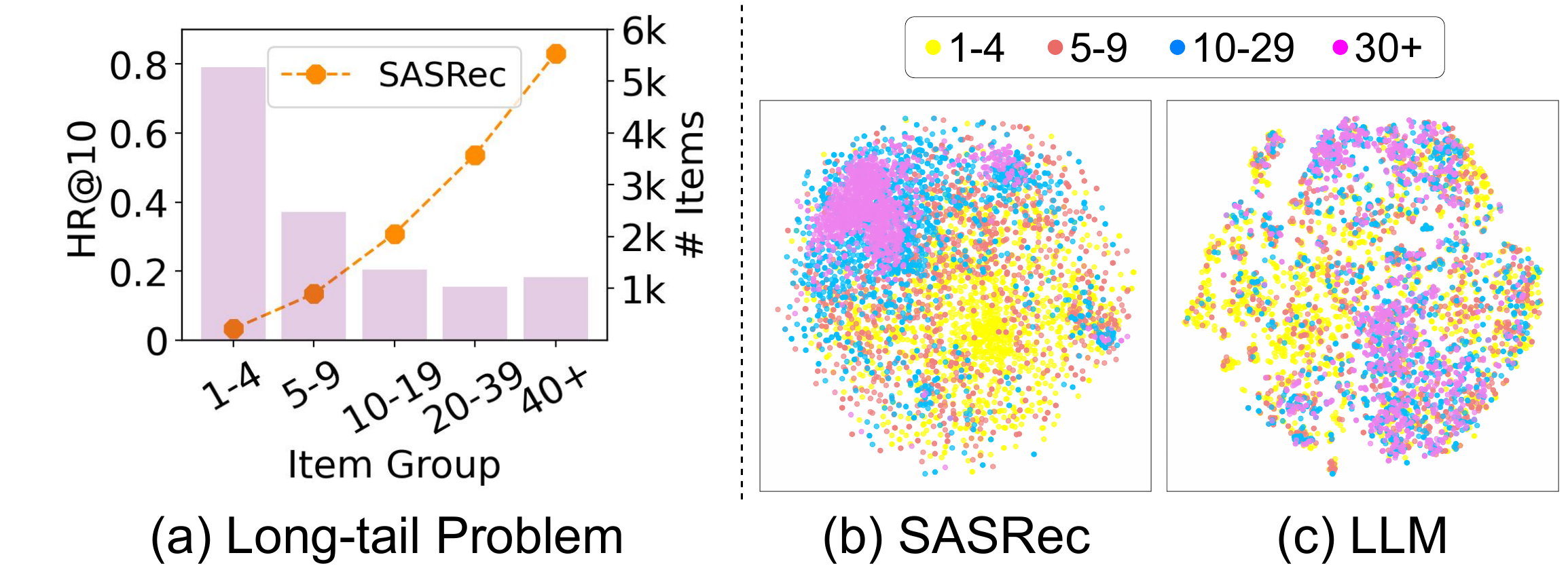}
\caption{The preliminary experiments are conducted for SASRec on the Yelp dataset. (a) The illustration for the long-tail problem. (b) The distribution of item embeddings of SASRec. (c) The distribution of LLM embeddings (input item texts and take out the last hidden state of LLaMA-7B).}
\label{fig:preliminary1}
\end{figure}

Although the accuracy of SRS has seen continuous improvement, the long-tail problem remains a critical challenge that can undermine the overall user experience.
To illustrate this issue, we trained a popular SRS model, SASRec, on the Yelp dataset and grouped the items based on their interaction frequencies. As depicted in Figure~\ref{fig:preliminary1}(a), the histogram reveals that the majority of items have fewer than $5$ records, while the corresponding line graph indicates their relatively low performance. 
This phenomenon highlights the difficulty in effectively recommending long-tail items, which can result in reduced serendipity for users and diminished profits for sellers.
Our analysis suggests that the long-tail problem in SRS primarily stems from the skewed distribution of item embeddings.
To further investigate this, we visualized the item embedding distribution of SASRec using t-SNE in Figure~\ref{fig:preliminary1}(b). 
The result confirms that the embeddings of low-popularity items (i.e., $1$-$4$) are sparsely distributed and distant from those of more popular items, indicating the poor quality of these embeddings. 
In contrast, the Large Language Model (LLM) shows promise in capturing semantic relationships between items through textual features, such as titles. Figure~\ref{fig:preliminary1}(c) shows the item embeddings generated by LLaMA~\cite{touvron2023llama} are more uniformly distributed, which motivates the development of an LLM-based generator for producing higher-quality embeddings.

\begin{figure}[!t]
\centering
\includegraphics[width=1\linewidth]{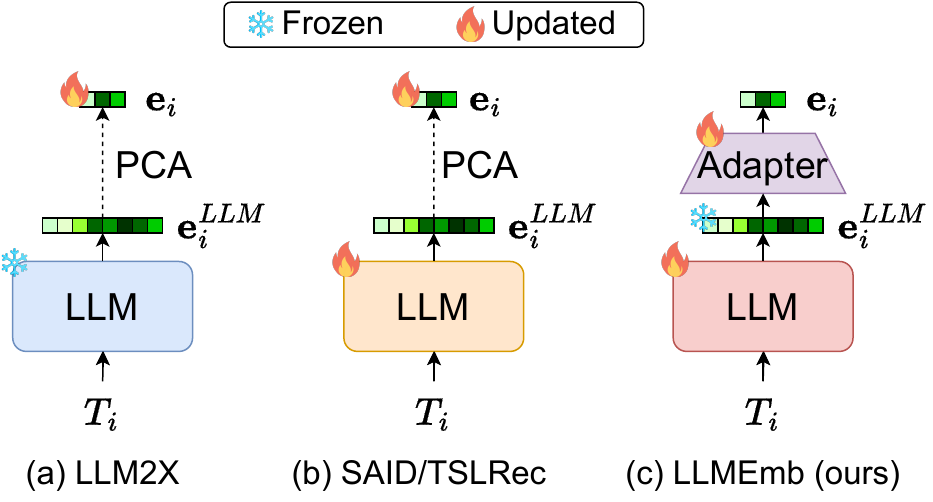}
\caption{The comparison between the existing LLM enhanced SRS methods and our LLMEmb.}
\label{fig:preliminary2}
\end{figure}

Some recent studies have explored the potential of leveraging LLM to enhance SRS~\cite{harte2023leveraging,hu2024enhancing,liu2024practice}. 
However, they encounter two significant challenges.
(i) \textbf{Semantic Gap}: LLM2X~\cite{harte2023leveraging} adopts the general-purpose LLM to generate item embeddings. 
While these embeddings can contain the semantics, they are not tailored to the recommendation field. 
In an effort to address this, methods like SAID~\cite{hu2024enhancing} and TSLRec~\cite{liu2024practice} propose fine-tuning open-sourced LLM to better align with recommendation tasks. 
However, these approaches remain confined to language modeling or category prediction, overlooking the crucial role of item attributes in distinguishing items within the recommendation field~\cite{hou2019explainable}.
(ii) \textbf{Semantic loss}: As shown in Figure~\ref{fig:preliminary2}, to further adapt the LLM embeddings to collaborative SRS models, existing methods reduce the dimension and update the embeddings with SRS models directly. 
However, drastic dimensionality reduction and continual training can result in a significant loss of the original semantic richness contained in LLM embeddings, thereby limiting their effectiveness, particularly for long-tail items.

To address the above challenges, we propose an LLM-based item embedding generator (\textbf{LLMEmb}) specified for SRS. The proposed LLMEmb involves a two-stage training. 
For the first stage, we design a Supervised Contrastive Fine-Tuning (SCFT) to bridge the semantic gap between general and recommendation domains. In detail, attribute-level data augmentation is designed to construct the training pairs for enhancing the distinguishing abilities of LLM. The fine-tuned LLM can derive recommendation-friendly embeddings.
The second stage, \ie Recommendation Adaptation Training (RAT), focuses on injecting the collaborative signals into LLM embeddings. To prevent semantic loss, we design a trainable adapter that allows for dimension transformation while keeping the LLM embeddings frozen. During inference, the generated embeddings can be cached into the embedding layer of SRS models, ensuring that no additional computational burden is introduced.
The contributions of this paper are concluded as follows:
\begin{itemize}[leftmargin=*]
    \item We design a novel LLM-based item embedding generator, which can help alleviate the long-tail problem for the sequential recommendation. 
    \item To fill the semantic gap between general and recommendation domains, we propose an attribute-level contrastive fine-tuning method. To avoid semantic loss, we fabricate a recommendation adaptation strategy. 
    \item We have conducted comprehensive experiments on three public datasets and verified the superior performance of LLMEmb combined with three popular SRS models. 
\end{itemize}

\section{Preliminary}

\textbf{Problem Definition}.
Let $v_i \in \mathcal{V}$ denotes the item in an item set, then the input sequence of user $u$ can be represented as $\mathcal{Q}^{(u)}=\{v_1^{(u)},v_2^{(u)},\dots,v_{N_u}^{(u)}\}$, which is ordered by timeline. $N_u$ is the length of the interaction sequence. 
For simplicity, we omit the user-specific superscript $(u)$ in subsequent notations. The task of recommending the next item can thus be formulated as:
\begin{equation}
    \text{arg} \max_{v_j \in \mathcal{V}} P(v_{N+1}=v_j|\mathcal{Q})
\end{equation}

\noindent \textbf{General SRS Framework}.
Most existing SRS models can be broadly concluded into a two-step framework known as \textbf{\textit{Embedding-Sequence}}. 
In the first step, the item identities are transformed into dense embeddings to represent them in a high-dimension space, capturing the collaborative relationships among items. 
Here, we denote item $i$ as $v_i$. The \textbf{\textit{Embedding}} procedure is formalized as $\mathbf{e}_i = {\rm Emb}(v_i)$.
${\rm Emb}(\cdot)$ denotes the \textbf{embedding function}, and the resulting $\mathbf{e}_i \in \mathbb{R}^{d}$ represents the high-dimensional embedding of the item $i$, with $d$ being the dimension size. 
After the first step, the input sequence is transformed into an embedding sequence $\bar{\mathcal{Q}}=\{\mathbf{e}_1,\mathbf{e}_2,\dots,\mathbf{e}_{N}\}$. 
The next step is the \textit{\textbf{Sequence}} procedure, which aims to extract the user preference from interaction histories. Thus, it absorbs $\bar{\mathcal{Q}}$ and outputs the representation of the user $\mathbf{u} \in \mathbb{R}^d$. The process can be represented as $\mathbf{u} = {\rm Seq}(\{\mathbf{e}_1,\mathbf{e}_2,\dots,\mathbf{e}_{N}\})$.
${\rm Seq}(\cdot)$ is the sequence modeling function, referred to as \textbf{SRS backbone} in this paper. Finally, the recommending probability of each item is calculated by taking the inner product of the user and item representations, \ie $P(v_{N+1}=v_i|\mathcal{Q})=\mathbf{u}^T \mathbf{e}_i$. 
Let $\hat{\mathbf{y}}$ denote the probability vector of all items. The framework is then optimized using a loss function, such as Binary Cross-Entropy calculated based on $\hat{\mathbf{y}}$.

To model user preferences more precisely, various neural architectures have been fabricated for \textbf{SRS backbone} ${\rm Seq}(\cdot)$, such as recurrent neural networks~\cite{cho2014learning} for GRU4Rec~\cite{hidasi2015session} and self-attention~\cite{vaswani2017attention} for SASRec~\cite{kang2018self}. However, the embedding function is often simply designed as a randomly initialized embedding layer and trained from scratch. In this paper, we focus on \textbf{utilizing the LLM to generate better embedding function}, \ie ${\rm Emb}(\cdot)$, which can be integrated into most SRS models.

\begin{figure*}[!t]
\centering
\includegraphics[width=0.9\linewidth]{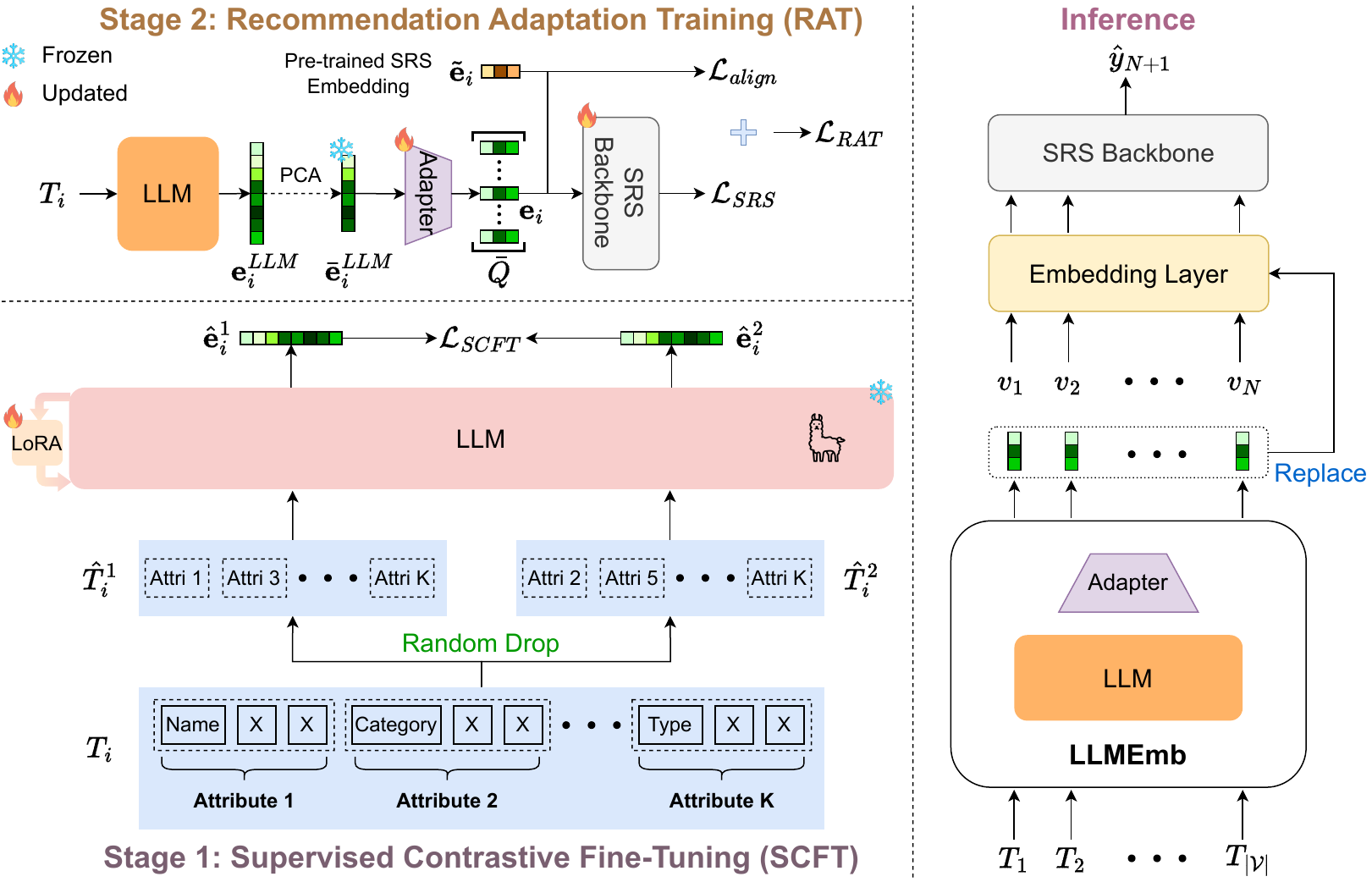}
\caption{The overview of the proposed LLMEmb.} 
\label{fig:overview}
\end{figure*}

\section{Method}

In this section, we will introduce the details of the proposed LLMEmb. Firstly, we will give an overview of our method. Then, the supervised contrastive fine-tuning will be addressed to illustrate how we fine-tune a general LLM into a recommendation-friendly one. Next, to further combine the collaborative signals with the LLM embeddings, we devise the recommendation adaptation training for LLMEmb. Finally, the training and inference process will be detailed.

\subsection{Overview}

Figure~\ref{fig:overview} shows the training and inference process of LLMEmb, composed of an LLM and an adapter.
For the training process, there are two stages specialized for the LLM and adapter, respectively.
In the first stage, known as \textbf{Supervised Contrastive Fine-tuning (SCFT)}, the objective is to fine-tune the general-purpose LLM to enhance its ability to distinguish items based on their various attributes. 
Specifically, the textual prompt of one item, composed of its attributes, will be augmented into two copies by randomly dropping a certain ratio of attributes. 
Then, we will fine-tune the LLM by contrasting its embedding of distinct items. After that, it can derive recommendation-friendly LLM embeddings, containing the semantic information of items. 
The second stage, termed \textbf{Recommendation Adaptation Training (RAT)}, involves training the adapter designed to transform the LLM embeddings into the final item embeddings.
These item embeddings are then fed into the SRS backbone and optimized using general recommendation loss.
During the inference phase, the LLMEmb will generate all item embeddings in advance. These precomputed embeddings substitute the original embedding layer of the SRS model.

\subsection{Supervised Contrastive Fine-Tuning (SCFT)}

The LLM has demonstrated exceptional semantic understanding capabilities across various natural language processing tasks~\cite{zhao2023survey,chang2024survey}, suggesting the potential to enhance SRS by extracting rich semantic information from the item's textual attributes.
However, most LLMs are trained for general purposes and may struggle to perceive subtle distinctions between items with varying attributes. 
To address this semantic gap, we design a supervised contrastive fine-tuning for the LLM (LLaMA in this paper). 
The key idea is enabling the LLM to distinguish between items by contrasting their attributes.



\noindent \textbf{\textit{Prompt Construction}}.
To encourage the LLM to understand the item from a semantic perspective, we construct textual prompts based on its attributes, \eg name, categories, and others.
The designed prompt consists of two parts. 
One is a domain-related instruction, denoted as $I$, to inform the LLM about the type of recommendation task. For example, the instruction can be ``\textit{The point of interest has the following attributes: }'' for a POI recommendation~\cite{long2023decentralized}. The other part includes all the attributes of the item, where each attribute is structured in the format ``\textit{$<$Attri$>$ is $<$Content$>$}''.
Here, \textit{$<$Attri$>$} and \textit{$<$Content$>$} will be replaced by the attribute name and actual attribute values. Let $A_j$ denote the atomic prompt for each attribute, then the prompt of item $i$ can be formulated as follows:
\begin{equation}
    T_i=[I, A_1, A_2, \dots, A_K]
\end{equation}
\noindent where $[\cdot]$ represents the concatenation operation for strings and $K$ is the number of attributes.

\noindent \textbf{\textit{Data Augmentation}}.
As previously discussed, our goal is to fine-tune the LLM to equip it with the capacity to distinguish the items with different attributes. 
Fundamentally, each item can be considered a negative sample relative to other items, as they represent distinct semantics within the recommendation. By fine-tuning the LLM to push the distance between different items, we improve the uniformity of semantic representations~\cite{ou2024prototypical}, which can subsequently enhance recommendation adaptability.
Then, to emphasize the fine-grained impact of item attributes, we propose to randomly drop a certain ratio of the item's attributes to get two copies of one item. These two copies serve as a pair of positive samples. Specifically, the augmentation process is as follows:
\begin{equation}
\begin{aligned}
    \hat{T}^1_i&=[I, {\rm RandomDrop}(\{A_j\}_{j=1}^K, r)] \\
    \hat{T}^2_i&=[I, {\rm RandomDrop}(\{A_j\}_{j=1}^K, r)]
\end{aligned}
\end{equation}
\noindent where ${\rm RandomDrop}(\cdot)$ denotes the operation of randomly dropping and $r$ is the ratio for dropping.

\noindent \textbf{\textit{Contrastive Fine-tuning}}.
Recent research has demonstrated that LLM can effectively generate high-quality embeddings for text, which are useful for tasks such as retrieval and matching~\cite{lee2024nv,wang2023improving}. 
Inspired by these works, we propose to utilize the LLM embeddings as the semantic representation of items. 
In detail, for each item $i$, we input prompt $T_i$ into the LLM and then average the corresponding word token embeddings from the final transformer layer to produce the LLM embedding, mark as $\mathbf{e}^{LLM}_i \in \mathbb{R}^{d_{token}}$. $d_{token}$ represents the dimension of token embedding in the LLM. 
We then apply in-batch contrastive learning~\cite{yang2023batchsampler} directly to these LLM embeddings. 
In detail, the augmented textual prompts, $\hat{T}^1_i$ and $\hat{T}^1_2$, are fed into the LLM,
producing the corresponding embeddings $\hat{\mathbf{e}}^{1}_i$ and $\hat{\mathbf{e}}^{2}_i$ for each item $i$. 
After that, the contrastive loss for one side augmentation can be expressed as follows:
\begin{equation} \label{eq:cl_llm}
    \mathcal{L}^1_{CL}=-\frac{1}{B} \sum_{i=1}^{B} \log \frac{\exp({\rm sim}(\hat{\mathbf{e}}^1_i, \hat{\mathbf{e}}^2_i) / \tau)}{\sum_{k=1}^B \mathbb{I}_{[i \neq k]} \exp({\rm sim}(\hat{\mathbf{e}}^1_i, \hat{\mathbf{e}}^2_k) / \tau)}
\end{equation}
\noindent where $\mathbb{I}_{[i \neq k]}\in \{0,1\}$ is an indicator function and $B$ is the batch size. ${\rm sim}(\cdot)$ is a similarity measuring function, which is the inner product in this paper. 
$\tau$ is a trainable temperature coefficient. In the same way, we can get the other side of contrastive loss $ \mathcal{L}^2_{CL}$ by exchanging the positions of $\hat{\mathbf{e}}^1$ and $\hat{\mathbf{e}}^2$ in the Equation~\eqref{eq:cl_llm}.
The final loss function used for fine-tuning the LLM is given by:
\begin{equation}
    \mathcal{L}_{SCFT} = \mathcal{L}^1_{CL} + \mathcal{L}^2_{CL}
\end{equation}

\subsection{Recommendation Adaptation Training (RAT)}
While the fine-tuned LLM can generate embeddings that better suit recommendation tasks, two key challenges remain when integrating these embeddings into SRS. 
The first challenge is the lack of collaborative signals, which are crucial for the effectiveness of SRS models~\cite{cai2021category}. 
The second challenge is dimension incompatibility, as the LLM embeddings often largely differ in size from the embeddings typically used in SRS models. 
To address these challenges, we introduce a Recommendation Adaptation Training (RAT) designed to transform LLM-generated embeddings into final item embeddings suitable for SRS models.
The RAT framework consists of three key components. The first component is \textbf{Embedding Transformation}, which integrates a trainable adapter to adjust the dimensionality of the LLM embeddings. 
The second component, \textbf{Adaptation}, involves injecting collaborative signals into the adapter by training it alongside the SRS backbone. 
Finally, \textbf{Collaborative Alignment} is devised to assist the optimization.

\noindent \textbf{\textit{Embedding Transformation}}.
Previous works~\cite{harte2023leveraging,hu2024enhancing} have proposed using PCA~\cite{pearson1901liii} to reduce the dimension of LLM embeddings, but they face semantic loss. 
To alleviate this problem, we propose a two-level transformation strategy. 
At the first level, we also apply PCA to reduce the embedding size, facilitating optimization~\cite{goodfellow2014qualitatively}. 
However, to preserve the semantics contained in LLM embeddings, we limit the reduction to an intermediate size (\eg $1536$), which remains significantly larger than the typical dimensionality of SRS embeddings (usually $128$).
This process can be formatted as $\mathbf{e}^{LLM} \stackrel{PCA}{\longrightarrow} \bar{\mathbf{e}}^{LLM}$, where $\mathbf{e}^{LLM}$ is the LLM embedding derived from the fine-tuned LLM, and $\bar{\mathbf{e}}^{LLM} \in \mathbb{R}^{d_m}$ denotes the downsized LLM embedding with $d_m$ being the intermediate size. 
Following this, we design an adapter to generate the final item embedding, ensuring compatibility with SRS models. For example, the final embedding of item $i$ can be computed as
\begin{equation}
    \mathbf{e}_i = \mathbf{W}_1(\mathbf{W}_2 \bar{\mathbf{e}}_i^{LLM})+\mathbf{b}_2)+\mathbf{b}_1
\end{equation}
\noindent where $\mathbf{W}_1 \in \mathbb{R}^{d \times \frac{d_m}{2}}, \mathbf{W}_2 \in \mathbb{R}^{{\frac{d_m}{2} \times d_m}}$ and $\mathbf{b}_1 \in \mathbb{R}^{d \times 1}, \mathbf{b}_2 \in \mathbb{R}^{\frac{d_m}{2} \times 1}$ are parameters of the two-layer adapter. By this transformation process, we can get the final item embeddings from LLM embeddings.

\noindent \textbf{\textit{Adaptation}}.
Although the semantic relationships captured by LLM embeddings can significantly benefit long-tail items, the incorporation of collaborative signals remains essential for effective recommendation tasks~\cite{cai2021category}. 
Thus, we design an adaptation process to train the derived embeddings. Specifically, we treat the LLM embeddings $\bar{\mathbf{e}}^{LLM}$, along with the proposed adapter, as the embedding function. These embeddings are then combined with an SRS backbone to complete the sequential recommendation process.
To learn collaborative signals, we update the randomly initialized SRS backbone and the adapter using the loss function specific to the corresponding SRS model, denoted as $\mathcal{L}_{SRS}$. For example, SASRec~\cite{kang2018self} adopts the Binary Cross-Entropy loss. It is worth noting that we freeze the parameters of the LLM embeddings $\bar{\mathbf{e}}^{LLM}$ during training, because the update of it will destroy the original semantic relationships. 
Consequently, during the RAT stage, only the parameters of the SRS backbone and the adapter are updated.

\noindent \textbf{\textit{Collaborative Alignment}}.
As mentioned earlier, only the adapter is trained to transform the semantic LLM embeddings into the final item embeddings. 
However, this approach may lead to overfitting, as only a small proportion of parameters (\ie those of the adapter) are updated.  
To mitigate this problem, we propose to align the derived item embeddings with the well-trained collaborative embeddings. 
Such an alignment will assist the optimization process by learning coarse collaborative relationships between items. Specifically, we first train an SRS model and take out its embedding layer. Let $\tilde{\mathbf{e}}_i$ denote item $i$'s embedding of the well-trained SRS model. Then, we design an in-batch contrastive loss to align $\mathbf{e}_i$ with $\tilde{\mathbf{e}}_i$:
\begin{equation}
    \mathcal{L}^1_{align}=-\frac{1}{S} \sum_{i=1}^{S} \log \frac{\exp({\rm sim}(\mathbf{e}_i, \tilde{\mathbf{e}}_i) / \gamma)}{\sum_{k=1}^S \mathbb{I}_{[i \neq k]} \exp({\rm sim}(\mathbf{e}_i, \tilde{\mathbf{e}}_k) / \gamma)}
\end{equation}
\noindent where $S$ and $\gamma$ denote the sum of sequence lengths of one batch and the temperature for contrastive learning, respectively. 
Similarly, we can compute the contrastive loss $\mathcal{L}^2_{align}$ that aligns $\tilde{\mathbf{e}}_i$ with $\mathbf{e}_i$. The sum of these two losses is denoted as $\mathcal{L}_{align}$, used for training the adapter and SRS backbone together with $\mathcal{L}_{SRS}$.

\subsection{Training and Inference}
In this section, we will detail the training and inference process. 

\noindent \textbf{\textit{Training}}. During the SCFT stage, we adopt the LoRA~\cite{hulora} technique to fine-tune the LLM, allowing us to save computational resources. 
Consequently, only the low-rank matrics $\{\mathbf{A}_i,\mathbf{B}_i\}_{i=1}^M$ are trained by $\mathcal{L}_{SCFT}$, where $M$ is number of layers accompanied by LoRA. In the RAT stage, the optimization process is formulated as:
\begin{equation}
    arg \min_{\Theta, \Phi} \mathcal{L}_{SRS} + \alpha \cdot \mathcal{L}_{align}
\end{equation}
\noindent where $\Theta$ represents the parameters of SRS backbone and $\Phi=\{\mathbf{W}_1,\mathbf{W}_2,\mathbf{b}_1,\mathbf{b}_2\}$ is the ones of the adapter.
The hyperparameter $\alpha$ controls the strength of the alignment.

\noindent \textbf{\textit{Inference}}. 
As previously described, a general SRS consists of an SRS backbone and an embedding function. 
During inference, the well-trained SRS backbone (\ie parameter $\Theta$) obtained from the RAT stage is used for the \textbf{\textit{Sequence}} procedure. 
For the embedding function, we generate LLM embeddings for all items using their textual prompts $T_i$ and then feed the dimension-reduced $\bar{\mathbf{e}}^{LLM}$ to the adapter. 
As a result, the final embeddings $\mathbf{e}_i \in \mathbb{R}^d$ are precomputed and cached in advance. 
These generated embeddings replace the weights of the embedding layer, effectively serving as the \textbf{\textit{Embedding}} component. 
In conclusion, this approach introduces no additional computational burden during inference compared to traditional SRS models.

\section{Experiment}
In this section, we will show the experimental results to respond to the following Research Questions (\textbf{RQ}). 

\begin{itemize}
    \item \textbf{RQ1}: How does the proposed LLMEmb perform, compared with LLM-based baselines? Can the LLMEmb enhance various SRS models?
    \item \textbf{RQ2}: Do all designs for LLMEmb take effect?
    \item \textbf{RQ3}: How do hyper-parameters affect the performance of our LLMEmb?
    \item \textbf{RQ4}: Can the proposed LLMEmb alleviate the long-tail problem in SRS?
    \item \textbf{RQ5}: Can LLMEmb correct embedding distributions?
\end{itemize}

\subsection{Experimental Settings}

\noindent \textbf{\textit{Dataset}}.
In the experiments, we adopt three real-world datasets for verification, \ie
Yelp, Amazon Beauty, and Amazon Fashion. Yelp includes amounts of check-in records, which can be used for point-of-interest recommendation. Amazon is collected from an e-commerce platform. Beauty and Fashion are two sub-categories of this dataset. We follow the preprocessing of the previous SRS works~\cite{kang2018self}.

\noindent \textbf{\textit{Sequential Recommendation Backbones}}.
Since the proposed LLMEmb is a model-agnostic method, it can be integrated with many SRS models. To verify the generality, we test our method and competing baselines on
GRU4Rec~\cite{hidasi2015session}, Bert4Rec~\cite{sun2019bert4rec}, and SASRec~\cite{kang2018self}. 

\noindent \textbf{\textit{Baselines}}.
To verify the effectiveness of our LLMEmb, we compare one state-of-the-art \textbf{Long-tail Sequential Recommendation} baseline, \ie MELT~\cite{kim2023melt}, and three up-to-date \textbf{LLM-enhanced Sequential Recommendation} baselines, including LLM2X~\cite{harte2023leveraging}, SAID~\cite{hu2024enhancing}, TSLRec~\cite{liu2024practice}. 

\noindent \textbf{\textit{Implementation Details}}.
All the experiments in this paper are conducted on an Intel Xeon Gold 6133 platform, equipped with Tesla V100 GPUs. The code is built on Python 3.9.5 with PyTorch 1.12.0. For a fair comparison, the foundation model used for baselines TSLRec and SAID, and our LLMEmb is LLaMA-7B~\cite{touvron2023llama}. 

\begin{table*}[t]
\centering
\tabcolsep=0.07cm   
\resizebox{\textwidth}{!}{
\begin{tabular}{ll|cccc|cccc|cccc}
\toprule[1pt]
\multirow{3}{*}{\textbf{Backbone}} & \multirow{3}{*}{\textbf{Model}} & \multicolumn{4}{c|}{\textbf{Yelp}} & \multicolumn{4}{c|}{\textbf{Fashion}} & \multicolumn{4}{c}{\textbf{Beauty}} \\ 
\cmidrule{3-14} 
&  & \multicolumn{2}{c}{\textbf{Overall}} & \multicolumn{2}{c|}{\textbf{Tail}} & \multicolumn{2}{c}{\textbf{Overall}} & \multicolumn{2}{c|}{\textbf{Tail}} & \multicolumn{2}{c}{\textbf{Overall}} & \multicolumn{2}{c}{\textbf{Tail}} \\ 
\cmidrule{3-14} 
&  & H@10 & N@10 & H@10 & N@10 & H@10 & N@10 & H@10 & N@10 & H@10 & N@10 & H@10 & N@10 \\ 
\midrule
\multirow{6}{*}{GRU4Rec} 
& - None & 0.4879 & 0.2751 & 0.0171 & 0.0059 & 0.4798 & 0.3809 & 0.0257 & 0.0101 & 0.3683 & 0.2276 & 0.0796 & 0.0567 \\
& - MELT & \underline{0.4985} & \underline{0.2825} & 0.0201 & 0.0079 & 0.4884 & 0.3975 & 0.0291 & 0.0112 & 0.3702 & 0.2161 & 0.0009 & 0.0003 \\
& - LLM2X & 0.4872 & 0.2749 & 0.0201 & 0.0072 & 0.4881 & 0.4100 & 0.0264 & 0.0109 & 0.4151 & \underline{0.2713} & 0.0896 & 0.0637 \\
& - SAID & 0.4891 & 0.2764 & 0.0180 & 0.0062 & \underline{0.4920} & \underline{0.4168} & \underline{0.0347} & \underline{0.0151} & \underline{0.4193} & 0.2621 & \underline{0.0936} & \underline{0.0661} \\
& - TSLRec & 0.4528 & 0.2509 & \underline{0.0255} & \underline{0.0095} & 0.4814 & 0.4042 & 0.0149 & 0.0071 & 0.3119 & 0.1865 & 0.0750 & 0.0474 \\
& - \textbf{LLMEmb} & \textbf{0.5270}* & \textbf{0.2980}* & \textbf{0.1116}* & \textbf{0.0471}* & \textbf{0.5062}* & \textbf{0.4329}* & \textbf{0.1046}* & \textbf{0.0477}* & \textbf{0.4445}* & \textbf{0.2726} & \textbf{0.3183}* & \textbf{0.1793}* \\ 
\midrule
\multirow{6}{*}{Bert4Rec} 
& - None & 0.5307 & 0.3035 & 0.0115 & 0.0044 & 0.4668 & 0.3613 & 0.0142 & 0.0067 & 0.3984 & 0.2367 & 0.0101 & 0.0038 \\
& - MELT & \underline{0.6206} & 0.3770 & 0.0429 & 0.0149 & 0.4897 & 0.3810 & 0.0059 & 0.0019 & 0.4716 & 0.2965 & 0.0709 & 0.0291 \\
& - LLM2X & 0.6199 & \underline{0.3781} & 0.0874 & 0.0330 & 0.5109 & \underline{0.4159} & 0.0377 & 0.0169 & 0.5029 & 0.3209 & 0.0927 & 0.0451 \\
& - SAID & 0.6156 & 0.3732 & \underline{0.0973} & 0.0382 & \underline{0.5135} & 0.4124 & \underline{0.0694} & \underline{0.0433} & \underline{0.5127} & \underline{0.3360} & \underline{0.1124} & \underline{0.0664} \\
& - TSLRec & 0.6069 & 0.3680 & 0.0969 & \underline{0.0388} & 0.5078 & 0.4143 & 0.0418 & 0.0182 & 0.4936 & 0.3178 & 0.1013 & 0.0589 \\
& - \textbf{LLMEmb} & \textbf{0.6294}* & \textbf{0.3881}* & \textbf{0.1876}* & \textbf{0.1094}* & \textbf{0.5244}* & \textbf{0.4238}* & \textbf{0.1485}* & \textbf{0.0764}* & \textbf{0.5247}* & \textbf{0.3485}* & \textbf{0.2430}* & \textbf{0.1224}* \\ 
\midrule
\multirow{6}{*}{SASRec} 
& - None & 0.5940 & 0.3597 & 0.1142 & 0.0495 & 0.4956 & 0.4429 & 0.0454 & 0.0235 & 0.4388 & 0.3030 & 0.0870 & 0.0649 \\
& - MELT & 0.6257 & 0.3791 & 0.1015 & 0.0371 & 0.4875 & 0.4150 & 0.0368 & 0.0144 & 0.4334 & 0.2775 & 0.0460 & 0.0172 \\
& - LLM2X & \underline{0.6415} & \underline{0.3997} & \underline{0.1760} & \underline{0.0789} & 0.5210 & 0.4486 & 0.0768 & 0.0473 & 0.5043 & 0.3319 & \underline{0.1608} & \underline{0.0940} \\
& - SAID & 0.6277 & 0.3841 & 0.1548 & 0.0669 & \underline{0.5316} & \underline{0.4619} & \underline{0.0901} & \underline{0.0540} & \underline{0.5097} & 0.3343 & 0.1549 & 0.0906 \\
& - TSLRec & 0.6152 & 0.3795 & 0.1383 & 0.0620 & 0.5125 & 0.4594 & 0.0652 & 0.0382 & 0.4977 & \underline{0.3366} & 0.1211 & 0.0789 \\
& - \textbf{LLMEmb} & \textbf{0.6647}* & \textbf{0.4113}* & \textbf{0.2951}* & \textbf{0.1456}* & \textbf{0.5521}* & \textbf{0.4730}* & \textbf{0.1513}* & \textbf{0.0826}* & \textbf{0.5277}* & \textbf{0.3460}* & \textbf{0.4194}* & \textbf{0.2595}* \\ 
\bottomrule[1pt]
\end{tabular}
}
\caption{The overall results of competing methods and \textbf{LLMEmb} on three datasets. The boldface refers to the highest score, and the underline indicates the best result of the baselines. ``\textbf{{\Large *}}'' indicates the statistically significant improvements (\ie two-sided t-test with $p<0.05$) over the best baseline.}
\label{tab:exp_overall}
\end{table*}

\noindent \textbf{\textit{Evaluation Metrics}}.
Following the previous works~\cite{sun2019bert4rec,kang2018self}, we adopt common used Top-$10$ \textit{Normalized Discounted Cumulative Gain} (\textbf{N@10}) and \textit{Hit Rate} (\textbf{H@10}) as the metrics. Each positive item in the test set will be paired with $100$ randomly sampled uninteracted items to calculate the metrics. Besides, for the robustness of the results, we repeatedly conduct each experiment three times with random seeds $42$, $43$, $44$ and report the average values in the following tables and figures.

\subsection{Overall Performance (RQ1)}

    To respond to the \textbf{RQ1}, we show the overall and long-tail performance on three datasets in Table~\ref{tab:exp_overall}. Specifically, according to the Pareto principle, we divide the items with the popularity ranked at the last $80\%$ into the \textbf{Tail} group. The results indicate that our LLMEmb can achieve superior performance compared with all competitors. Especially, our method benefits the long-tail items with a large margin. For a more detailed analysis, we find that LLM-based methods often outperform MELT, a collaborative method for the long-tail problem. Such a phenomenon verifies the effectiveness of introducing semantics by the LLM. Comparing these three LLM-based methods, TSLRec often lags behind, because it only adopts the identities instead of textual information of items when using the LLM. Though SAID and LLM2X can also bring a large performance elevation to all SRS models, they are still inferior to LLMEmb, especially for the long-tail items. This comparison indicates our LLMEmb can better maintain the semantic relationship in the original LLM embeddings. In conclusion, due to our design of LLM fine-tuning and recommendation adaptation, the proposed LLMEmb can enhance the three SRS models consistently.

\begin{table}[!t]
\centering
\tabcolsep=0.1cm   
\resizebox{\linewidth}{!}{
\begin{tabular}{cc|cccc}
\toprule[1pt]
\multirow{2}{*}{\textbf{Dataset}} & \multirow{2}{*}{\textbf{Model}} & \multicolumn{2}{c}{\textbf{Overall}} & \multicolumn{2}{c}{\textbf{Tail}} \\ 
\cmidrule{3-6} 
 &  & H@10 & N@10 & H@10 & N@10 \\ 
 \midrule
\multirow{5}{*}{\textbf{Yelp}} 
& \textbf{LLMEmb} & \textbf{0.6647} & \textbf{0.4113} & \textbf{0.2951} & \textbf{0.1456} \\
& - \textit{w/o} SCFT & 0.6538 & 0.4031 & 0.2474 & 0.1218 \\ 
& - \textit{w/o} adapter & 0.6414 & 0.3968 & 0.2196 & 0.1055 \\
& - \textit{w/o} freeze & 0.6257 & 0.3800 & 0.1710 & 0.0740 \\
& - \textit{w/o} align & 0.6598 & 0.4060 & 0.2793 & 0.1310 \\
\bottomrule[1pt]
\end{tabular}
}
\caption{The ablation study conducted on the Yelp dataset and based on the SASRec backbone. The boldface refers to the highest score.}
\label{tab:exp_ablation}
\end{table}

\begin{figure}[!t]
\centering
\includegraphics[width=1\linewidth]{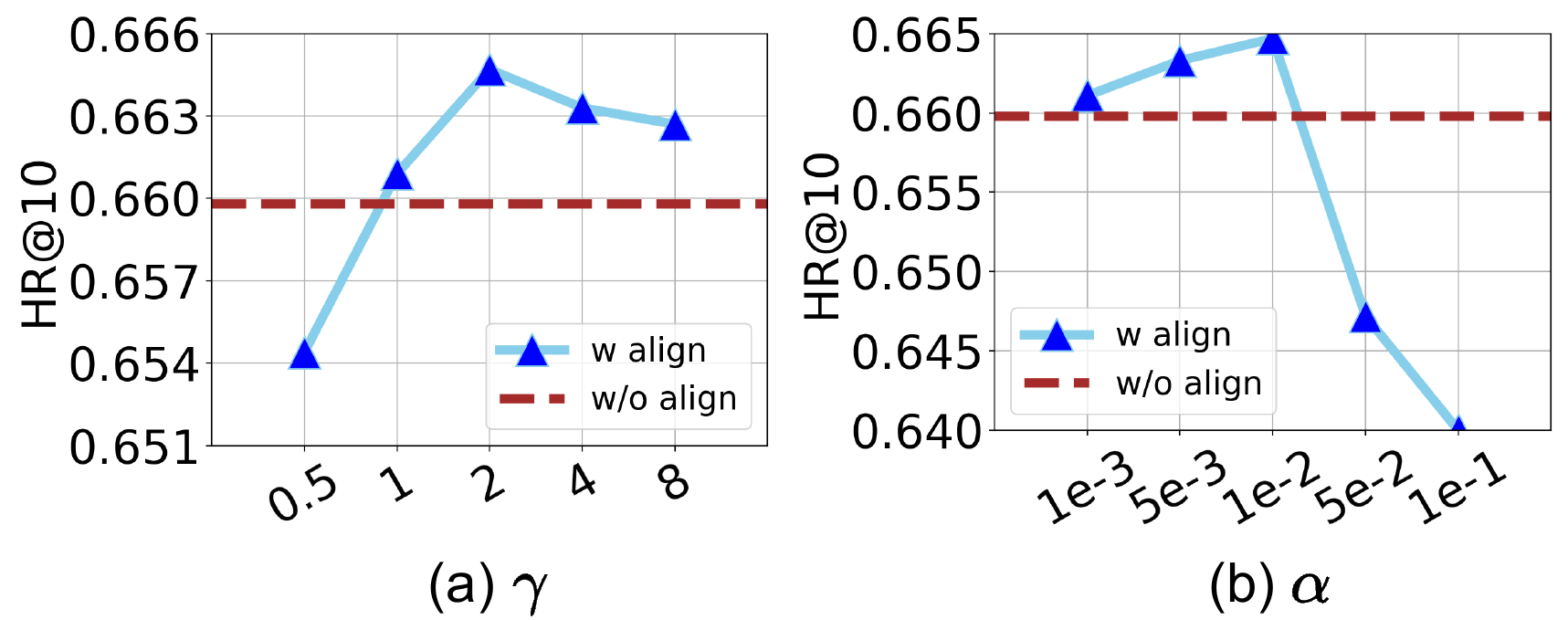}
\caption{The results of experiments for the temperature $\gamma$ and the weight $\alpha$ of alignment loss based on the Yelp dataset and SASRec backbone.}
\label{fig:exp_align}
\end{figure} 

\subsection{Ablation Study (RQ2)}

    For \textbf{RQ2}, we have conducted the ablation study and show the results in Table~\ref{tab:exp_ablation}. To investigate the effect of the proposed SCFT, we evaluate adopting the LLaMA without fine-tuning to derive LLM embeddings, denoted as \textit{w/o SCFT}. The performance of this variant drops under both overall and tail metrics, highlighting the necessity to fill the \textbf{semantic gap} between general LLM and recommendation tasks. Then, we evaluate three variants to verify the designs of RAT. \textit{w/o adapter} means removing the trainable adapter directly, which shows sub-optimal performance. It indicates the effectiveness of transformation. The variant without freezing the LLM embeddings during training, marked as \textit{w/o freeze}, severely harms the performance, suggesting the optimization difficulty in training large-size embedding. \textit{w/o align} eliminating the alignment loss directly illustrates the effectiveness of the collaborative alignment by the performance decrease.

\subsection{Hyper-parameter Analysis (RQ3)}

    The temperature $\gamma$ and scale $\alpha$ of the collaborative loss are two vital hyper-parameters in our LLMEmb. We show trends with their changes in Figure~\ref{fig:exp_align} to answer the \textbf{RQ3}. As the temperature $\gamma$ changes from $0.5$ to $8$, the overall HR@10 values of LLMEmb rise first and drop then. The reason lies in that the proper uniformity brought by contrastive learning can assist the optimization. In terms of the scale, the performance gets elevated with $\alpha$ rise from $1e^{-3}$ to $1e^{-2}$, which indicates the effectiveness of the designed alignment. However, larger $\alpha$ downgrades the performance, because the higher intensity of contrastive loss will lead to a convergence dilemma.
    Besides, the intermediate size $d_m$ is also important to the designed adapter. 

\subsection{Group Analysis (RQ4)}

\begin{figure}[!t]
\centering
\includegraphics[width=0.7\linewidth]{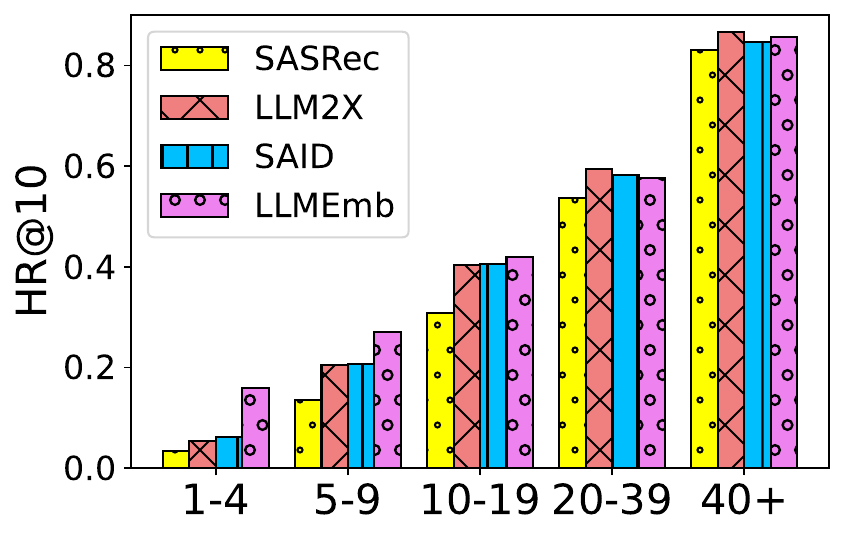}
\caption{The experimental results of group analysis based on Yelp dataset and SASRec backbone.}
\label{fig:exp_group}
\end{figure} 

    To explore the long-tail problem more carefully and answer the \textbf{RQ4}, we cluster the items by their popularity in $5$ groups and show the results in Figure~\ref{fig:exp_group}. Observing the figure, we find that the LLM-based methods can benefit the items with any popularity because of the semantic relationships. Compared with LLM2X and SAID, our LLMEmb brings more performance elevation to long-tail items, especially for the $1$-$4$ group. Such a phenomenon validates that our method can better integrate semantics from the LLM into recommendation. However, the LLMEmb underperforms LLM2X for those popular items (\eg $40+$ group) slightly, indicating a seesaw problem between popular and long-tail items. 

\subsection{Visualization (RQ5)}

\begin{figure}[!t]
\centering
\includegraphics[width=1\linewidth]{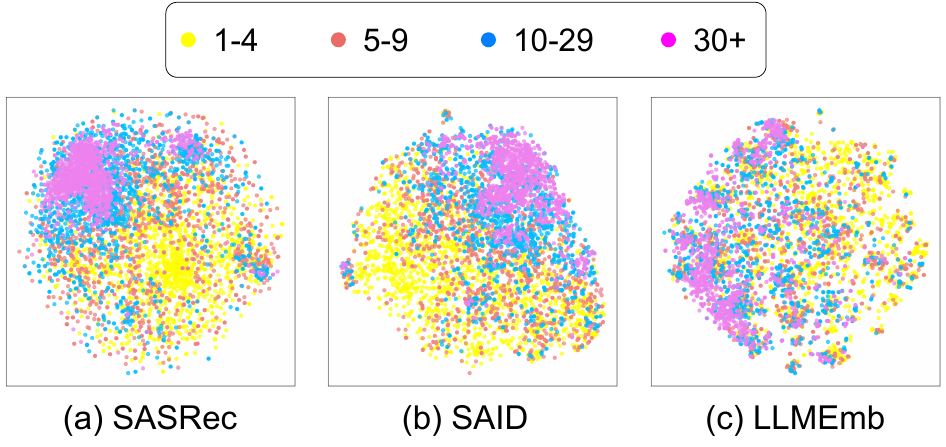}
\caption{The visualization of embeddings. The LLMEmb and baselines are based on SASRec and the Yelp dataset.}
\label{fig:exp_tsne}
\end{figure}

    To investigate whether the proposed LLMEmb can correct the skewed distribution of the embeddings, we visualize the distributions by t-SNE in Figure~\ref{fig:exp_tsne}. The figure shows that SAID can get a more even distribution by introducing the LLM embeddings. However, it is still congregated by the item's popularity due to the \textbf{semantic loss} issue. In contrast, our LLMEmb gets better embeddings, which are distributed more uniformly. The results respond to \textbf{RQ5} and reveal the superiority of our LLMEmb intrinsically. 

\section{Related Works}

\textbf{Sequential Recommendation}. 
The sequential recommendation aims to capture the user's preference from his or her historical interactions and then predict the next most possible item~\cite{liu2023disentangling,liu2023diffusion,liu2024bidirectional,li2023strec,liu2024sequential,liu2023linrec,li2023automlp,liang2023mmmlp,liu2023multi,liu2024multimodal,wang2023plate}. Many existing SRS works focus on fabricating the neural architecture to get the preference and dynamics more accurately. For example, Caser~\cite{tang2018personalized} adopts CNN for sequence modeling, while SASRec~\cite{kang2018self} firstly integrates self-attention~\cite{vaswani2017attention} layers. Later, for higher efficiency, some research studies~\cite{zhou2022filter} propose the MLP-based structure.
On the other hand, the loss function for training SRS models has also been highlighted in recent years. Bert4Rec~\cite{sun2019bert4rec} propose the cloze task to derive the training loss,
while CLS4Rec~\cite{liu2021contrastive} further designs the contrastive loss for training the SRS models.
However, most existing works have ignored the importance of item embeddings, which suffer from skewed distribution. In this paper, we propose an LLM-based method to construct better embeddings.


\noindent \textbf{Large Language Model for Recommendation}. 
Many efforts have been made to utilize the powerful LLM for recommendation~\cite{liu2024leader,liu2024llmesr,liu2024llmers}. A branch of research studies proposes to utilize the LLM for recommendation directly. For instance, TALLRec~\cite{bao2023tallrec} designs the textual prompt for recommendation tasks, which motivates the LLM to generate the predicted item name. 
Besides, to combine the collaborative signals into the LLM, E4SRec~\cite{li2023e4srec} and LLaRA~\cite{liao2023llara} design a trainable adapter to project the pre-trained item embeddings to language space and impose parameter-efficient fine-tuning~\cite{liu2024moelora}. Despite the brilliant performance of these models, the direct utilization of the LLM is resource-consuming, which is intolerant to real-time recommendation. For this issue, LLM2X~\cite{harte2023leveraging}, SAID~\cite{hu2024enhancing} and TSLRec~\cite{liu2024practice} propose to adopt the LLM embeddings to enhance SRS models. 
However, they still face semantic gap and loss, leading to sub-optimal performance.

\section{Conclusion}

In this paper, we propose a novel LLM-based generator, \ie LLMEmb, to derive item embeddings for the sequential recommendation. 
Specifically, to equip the LLM with the capacity to identify the items for recommendation tasks, we devise a supervised contrastive fine-tuning.  
Then, to avoid semantic loss and inject collaborative signals, we propose the recommendation adaptation training to update a trainable adapter. In the end, the well-trained LLM and adapter constitute the LLMEmb and can generate the final item embeddings. We conduct experiments on three real-world datasets and verify the effectiveness of LLMEmb.

\section{Acknowledgements}
    This research was partially supported by 
    National Key Research and Development Program of China (2022ZD0117100), National Natural Science Foundation of China (No.62192781, No.62177038, No.62293551, No.62277042, No.62137002, No.61721002, No.61937001, No.62377038), Project of China Knowledge Centre for Engineering Science and Technology, ``LENOVO-XJTU'' Intelligent Industry Joint Laboratory Project, 
    Research Impact Fund (No.R1015-23), APRC - CityU New Research Initiatives (No.9610565, Start-up Grant for New Faculty of CityU), CityU - HKIDS Early Career Research Grant (No.9360163), Hong Kong ITC Innovation and Technology Fund Midstream Research Programme for Universities Project (No.ITS/034/22MS), Hong Kong Environmental and Conservation Fund (No. 88/2022), and SIRG - CityU Strategic Interdisciplinary Research Grant (No.7020046), Tencent (CCF-Tencent Open Fund, Tencent Rhino-Bird Focused Research Program).  


\bibliography{main}


\end{document}